\documentclass[twocolumn,preprintnumbers,superscriptaddress,nofootinbib,aps,prd,floatfix]{revtex4}

\usepackage[utf8]{inputenc}

\usepackage{footmisc,enumerate}
\usepackage{multirow}
\usepackage{subfigure}
\usepackage{amsmath}
\usepackage{graphicx}
\usepackage{placeins}
\usepackage{xspace,slashed}
\usepackage{hyperref}
\hypersetup{colorlinks=true, citecolor=black, urlcolor=blue, linkcolor=blue}
\usepackage[normalem]{ulem}

\newcommand{\Eq}[1]{Eq.~\eqref{#1}}

\newcommand{\vev}{\ensuremath{v}\xspace}

\begin{document}

\newcommand{\og}{\ensuremath{\tilde{O}_g}\xspace}
\newcommand{\ot}{\ensuremath{\tilde{O}_t}\xspace}

\preprint{QMUL-PH-19-19}

\title{Effective field theory and scalar extensions of the top quark
  sector}

\begin{abstract}
Effective field theory (EFT) approaches are widely used at the LHC,
such that it is important to study their validity, and ease of
matching to specific new physics models. In this paper, we consider an
extension of the SM in which a top quark couples to a new heavy
scalar. We find the dimension six operators generated by this theory
at low energy, and match the EFT to the full theory up to NLO
precision in the simplified model coupling. We then examine the range
of validity of the EFT description in top pair production, finding
excellent validity even if the scalar mass is only slightly above LHC
energies, provided NLO corrections are included. In the absence of the
latter, the LO EFT overestimates kinematic distributions, such that
over-optimistic constraints on BSM contributions are obtained. We next
examine the constraints on the EFT and full models that are expected
to be obtained from both top pair and four top production at the LHC,
finding for low scalar masses that both processes show similar
exclusion power. However, for larger masses, estimated LHC
uncertainties push constraints into the non-perturbative regime, where
the full model is difficult to analyse, and thus not perturbatively
matchable to the EFT. This highlights the necessity to improve
uncertainties of SM hypotheses in top final states.
\end{abstract}

\author{Christoph Englert} \email{christoph.englert@glasgow.ac.uk}
\affiliation{SUPA, School of Physics and Astronomy, University of Glasgow, Glasgow G12 8QQ, UK\\[0.1cm]}
\author{Peter Galler} \email{peter.galler@glasgow.ac.uk}
\affiliation{SUPA, School of Physics and Astronomy, University of Glasgow, Glasgow G12 8QQ, UK\\[0.1cm]}
\author{Chris D. White} \email{christopher.white@qmul.ac.uk}
\affiliation{Centre for Research in String Theory, School of Physics and Astronomy, Queen Mary University of London, 327 Mile End Road,
London E1 4NS, UK\\[0.1cm]}


\maketitle

\section{Introduction}
\label{sec:intro}

The potential discovery of new physics beyond the Standard Model (BSM)
remains one of the principal motivations of contemporary high energy
physics research, in both theory and experiment. Much attention
focuses on the top quark and its antiparticle, given that these are
the heaviest particles in the SM, whose behaviour is thus likely to be
particularly sensitive to BSM effects. Furthermore, they are of
fundamental importance when discussing the naturalness (or otherwise)
of the electroweak symmetry breaking scale, such that typical BSM
scenarios necessarily involve modifications of the top sector. Their
effects may then be easier to investigate experimentally than purely
electroweak processes (e.g. Higgs production), owing to the relatively
large production cross sections of top particles at the Large Hadron
Collider (LHC) (see e.g. Ref.~\cite{Kroninger:2015oma} for a recent review). 

As is well-known there are, broadly speaking, two main ways to
investigate possible new physics. The first is to assume a particular
BSM theory, and to look for associated signatures, such as the
resonant production of new particles e.g. decaying to a top pair. This
is necessarily highly model-dependent, and lack of convincing new
physics signatures at the LHC to date instead motivates the use of
model-independent approaches. Chief amongst these is perhaps effective
field theory (EFT), in which one considers the SM Lagrangian to be the
leading term in an expansion in gauge-invariant higher-dimensional
operators. One may then extend the SM Lagrangian by these higher
dimensional terms, where dimension six is the state of the
art~\cite{Weinberg:1978kz,Buchmuller:1985jz,Burges:1983zg,Leung:1984ni,Hagiwara:1986vm,Grzadkowski:2010es,Dedes:2017zog}
(for a review see~\cite{Brivio:2017vri}). Each correction to the SM is
suppressed by one or more inverse powers of the new physics scale, and
thus such a framework is only applicable if this scale (e.g. a new
particle mass) exceeds the typical energy scales that are probed
in a particular collider of interest.

An intermediate approach between EFT and concrete UV scenarios is
represented by so-called simplified models (for reviews
see~\cite{Alwall:2008ag,Alves:2011wf,Abdallah:2015ter}), which aim to
reproduce a broad class of kinematic properties of the full UV
theories parametrised by only a few additional propagating degrees of
freedom and their couplings. The relation to the EFT approach is that
certain classes of operators have then been resummed to arbitrarily
high mass dimension, and certain extensions of the SM might be
particularly generic or well-motivated, e.g. Higgs mixing models.
Furthermore, experimental collaborations will in any case investigate
large classes of simplified models. It is then instructive, for
particular examples, to compare the two techniques, particularly with
regards to how constraints from the approaches to new physics compare
with each other in different kinematic regions.

In this paper, we perform a case study of this idea in the top quark
sector, which has been the subject of a number of recent EFT
studies~\cite{Buckley:2015nca,Rosello:2015sck,Buckley:2015lku,Zhang:2016omx,Bylund:2016phk,Castro:2016jjv,Barducci:2017ddn,Englert:2017dev,Schulze:2016qas,Birman:2016jhg,Cirigliano:2016nyn,Englert:2016aei,Maltoni:2016yxb,Zhang:2017mls,Etesami:2017ufk,AguilarSaavedra:2018nen,Malekhosseini:2018fgp,Degrande:2018fog,Jueid:2018wnj,DHondt:2018cww,Durieux:2018tev,deBeurs:2018pvs,Englert:2018byk,Chala:2018agk,Hartland:2019bjb,Durieux:2019rbz,Moutafis:2019wbp,Liu:2019wmi,Boughezal:2019xpp,Neumann:2019kvk}. We
will consider a particular model, in which the SM is supplemented by
an additional scalar, whose behaviour is parametrised entirely by its
mass and couplings. We will calculate the top pair production
cross-section including the effects of this new particle up to
one-loop order, showing explicitly which dimension six SM EFT operators
are generated upon taking the mass to be asymptotically
large. Matching of the EFT (see also \cite{delAguila:2016zcb,Zhang:2016pja,Ellis:2016enq,Henning:2016lyp,Ellis:2017jns,Summ:2018oko,Kramer:2019fwz} for generic approaches) to the full theory can be performed at
(next-to) leading order ((N)LO) in the coupling space of the latter,
so that we have potentially three different descriptions of the new
physics: (i) the LO EFT description; (ii) the NLO EFT description, in
which more effective operators are generated; (iii) the full
simplified model. We can then examine the validity of each approach,
and the ease of matching EFT constraints to the full theory.

We will first focus on top quark pair production, demonstrating
explicitly that an EFT description can provide an excellent
approximation to the full model, as expected. However, we will see
that NLO corrections in the EFT approach are particularly important,
and that a na\"{i}ve LO approach tends to overestimate kinematic
distributions, such that its (invalid) application would lead to
over-optimistic constraints on new physics.

The operators examined in this paper also affect four top
production~\cite{Cao:2016wib,Zhang:2017mls,Englert:2019zmt}, which is
actively being searched for by both the ATLAS~\cite{Aaboud:2018jsj}
and CMS~\cite{Sirunyan:2017roi,Sirunyan:2019nxl} collaborations. We
examine the projected constraints on this process (and top pair
production) that are expected to be obtained after the high luminosity
LHC (HL-LHC)
upgrade~\cite{CMS:1900mtx,ATL-PHYS-PUB-2018-047,Azzi:2019yne}, and
convert these into constraints on the parameter space of the new physics model.
We will see that constraints from four top production are competitive with top
pair production, suggesting that the two processes would have roughly
comparable weights in a global EFT fit. However, the extrapolated uncertainties
from both top pair and four top production lead to constraints that probe
parameter space regions in which the full theory is non-perturbative. For large
scalar masses, the width of the scalar resonance increases, such that no
meaningful constraint on the coupling is obtained in the full theory. Thus,
whilst constraints in the EFT description remain in principle valid and are
possible, it becomes impossible to match the EFT description to the full theory
of new physics, given that perturbative computations in the latter are not
obtainable.

The model considered here has been widely-studied in a
  number of different new physics scenarios. Thus, we hope that our
results provide a useful case study for the application of EFT at the
LHC, which will inform pragmatic discussions about how to apply this
technique going forwards, and what can be learned (or otherwise) about
specific UV completions. The structure of our paper is as follows. In
Sec.~\ref{sec:simp} we introduce the simplified model (of an
additional scalar particle) that we are considering, and calculate the
corrections to top pair production up to NLO. We furthermore explain
how the EFT description is obtained at low energy (relative to the
scalar mass). In section~\ref{sec:validity}, we present numerical
results for the top invariant mass distribution, and demonstrate the
validity of the EFT description, even at LO, when the scalar mass is
asymptotically large. We then quantify the mass regime in which the
NLO-matched EFT description is a good approximation of the full
theory. In section~\ref{sec:results}, we examine the projected
uncertainties on top pair and four top production at the LHC, and
examine the constraints obtained in the EFT at (N)LO, as well as the
full theory. Finally, in section~\ref{sec:discussion}, we discuss our
results and conclude.

\begin{figure*}[!t]
\centering
\includegraphics[width=0.65\textwidth]{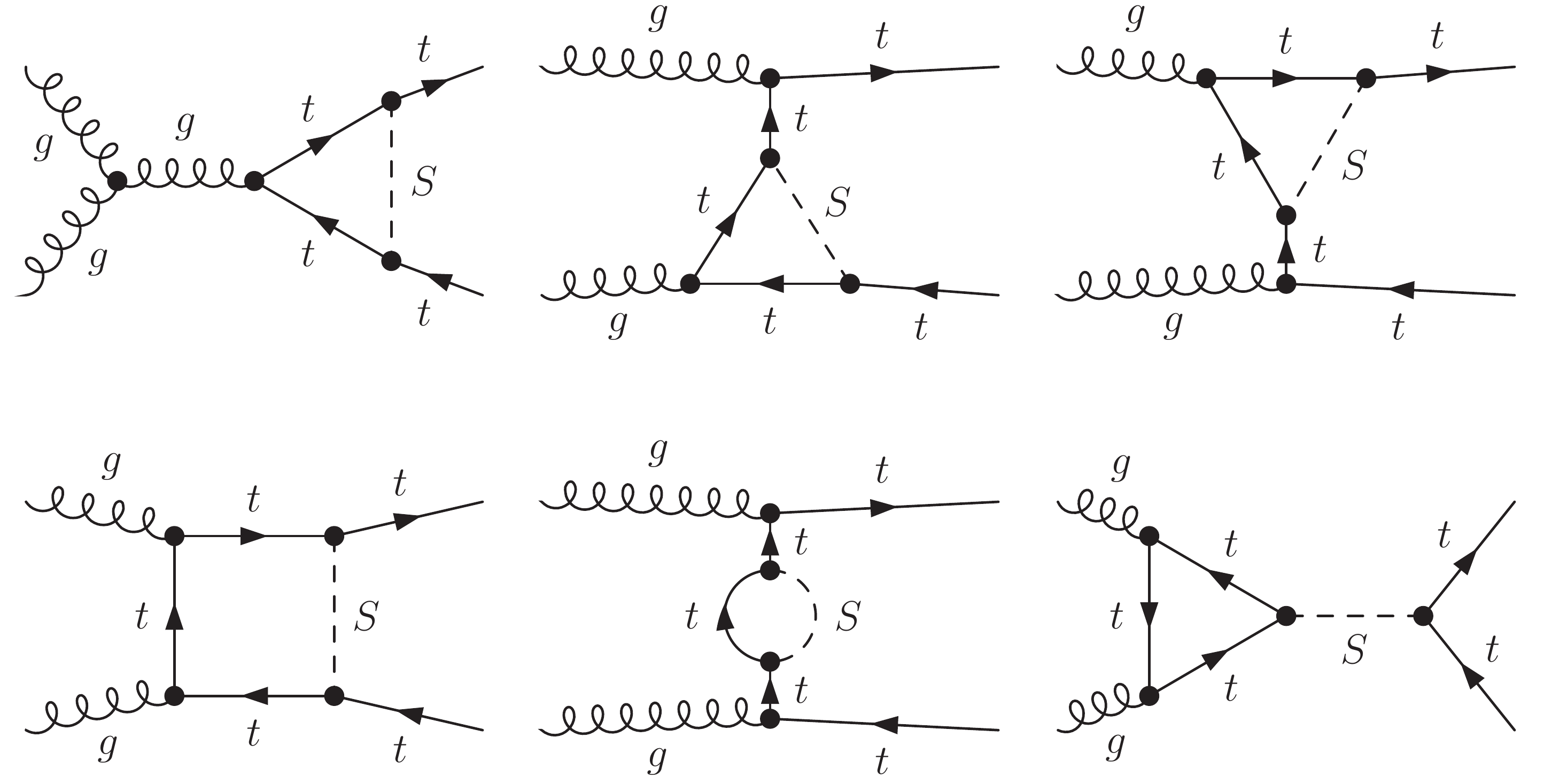}
\caption{Representative one-loop Feynman diagram contributions to $gg\to t\bar t$ arising in the simplified model of eq.~\eqref{eq:lagsimp}.\label{fig:full}}
\end{figure*}

\section{A Simplified Model and its EFT limit}
\label{sec:simp}
In this work, we consider a simplified model (similar to Ref.~\cite{Arina:2016cqj}) with dominant couplings
to the top quark
\begin{equation}
\label{eq:lagsimp}
{\cal{L}}_{{\text{BSM}}} ={1\over 2} \partial_\mu S \partial^\mu S - {1\over 2} m_S^2 S^2 - \left( c_S \bar t_L t_R S + {\text{h.c.}} \right)
\end{equation}
where $S$ is a scalar field of mass $m_S$.\footnote{Similar frameworks
have been considered in FCNC studies, e.g.~\cite{Banerjee:2018fsx}.} Provided the latter is
greater than $2m_t$, where $m_t$ is the top mass, the scalar $S$ may
directly decay into (anti)-top pairs, with corresponding width
\begin{equation}
\label{eq:partialtt}
\Gamma(S\to t\bar t) = {3 c_S^2 m_S\over 8\pi} \sqrt{1-{4m_t^2\over m_S^2}} \equiv c_S^2  \, \tilde{\Gamma} \,.
\end{equation}
Further contributions to the width arise from the fact that $S$ can
couple to gluons and photons via a top quark loop, analogously to the
SM Higgs boson. Although we include the loop-induced decays for
completeness, the prompt decay $S\to t \bar t$ dominates over the entire considered mass range. 
\begin{figure*}[!t]
\centering
\includegraphics[width=0.63\textwidth]{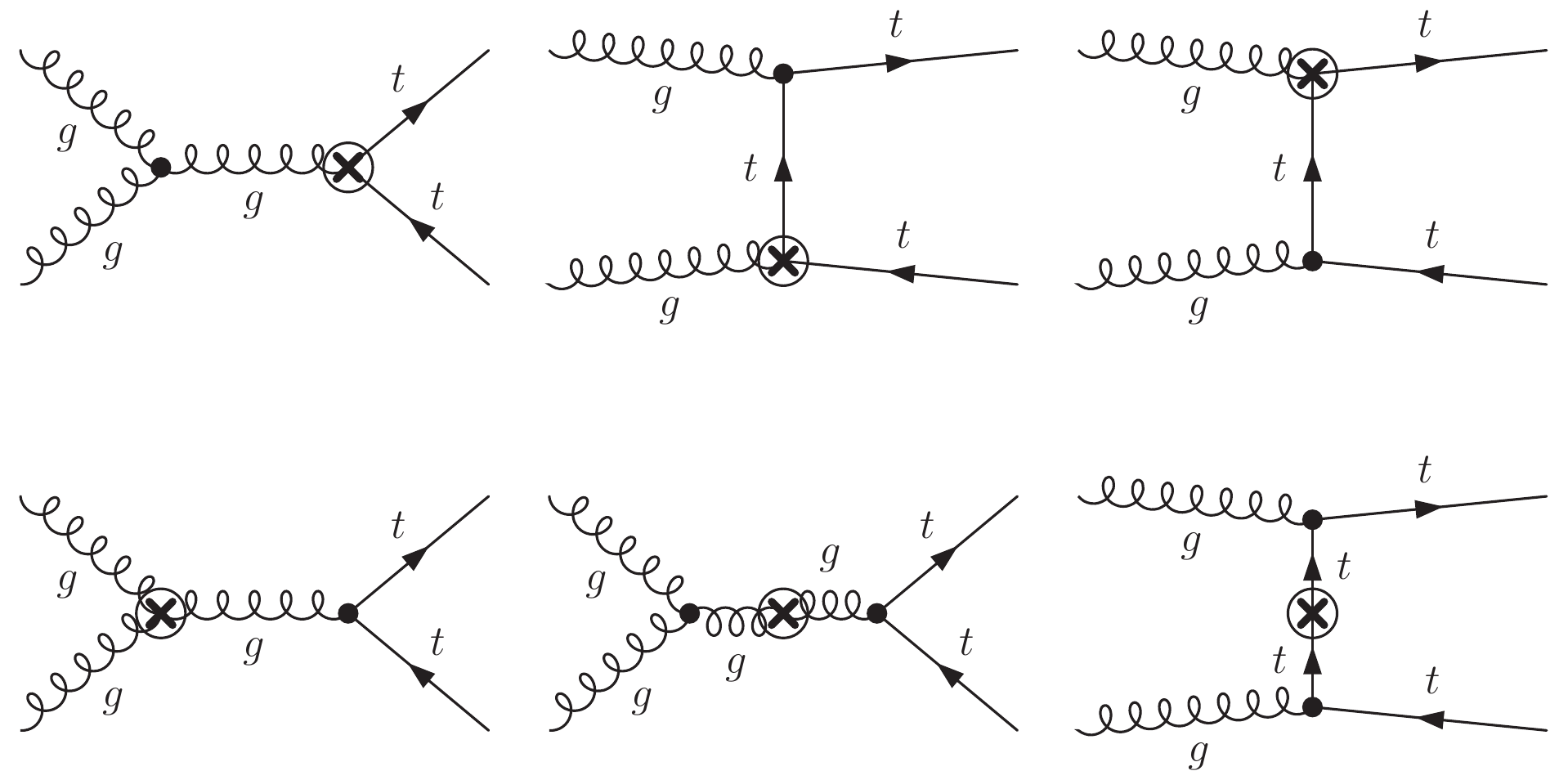}
\caption{\label{fig:effct} Representative counter term contributions to $gg\to t\bar t$. }
\end{figure*}

Our aim in this paper is to compare an EFT description of the theory
of eq.~\eqref{eq:lagsimp} at low energy, with the full theory, in
order to assess the validity and interpretation of the former. To this
end, let us consider how this theory leads to corrections to top pair
production up to NLO in the coupling of the scalar i.e. up to
and including ${\cal{O}}(c_S^2)$. Comparison with the EFT will then
allow us to match the two descriptions. Representative diagrams
contributing to the gluon-induced process $gg\rightarrow t\bar{t}$ are
shown in fig.~\ref{fig:full}, where we do not consider SM electroweak
contributions~\cite{Czakon:2017wor} (see
also~\cite{Vryonidou:2018eyv,Maltoni:2019aot}). In the SM, for heavy
Higgs bosons, it is known that the Higgs signal (with a large QCD $K$
factor~\cite{Anastasiou:2015ema,Anastasiou:2016cez}) has sizeable
interference effects with the QCD continuum in $gg\to t\bar
t$~\cite{Gaemers:1984sj,Dicus:1994bm,Bernreuther:1998qv}. This
influences exclusion constraints, but is also a viable source for new
physics beyond the
SM~\cite{Frederix:2007gi,Barger:2006hm,Craig:2015jba,Englert:2017dev,Bernreuther:2015fts,Hespel:2016qaf,Carena:2016npr,BuarqueFranzosi:2017qlm,Basler:2018dac,Djouadi:2019cbm,Kauer:2019qei}. The
predominant focus of previous work was therefore devoted to isolating
the resonance shape and cross section, which is not our focus
here. Note, however, that loop effects and their relation to (Higgs) effective field
theory were first discussed
in~\cite{Brehmer:2015rna,Franzosi:2015osa,Freitas:2016iwx,BuarqueFranzosi:2017jrj}.

For our analysis, we implement the leading order, virtual and counter
term (fig.~\ref{fig:effct}) contributions for $q\bar q , gg \to t\bar
t$ production at ${\cal{O}}(c_S^2)$ in a modified version of
{\sc{Vbfnlo}}~\cite{Arnold:2008rz,Arnold:2011wj,Arnold:2012xn,Baglio:2014uba}
which links
{\sc{FormCalc/LoopTools}}~\cite{Hahn:1999mt,Hahn:2001rv}. Various
analytical comparisons against alternative calculations as well as
numerical cross checks of leading order amplitudes have been performed
using {\sc{MadGraph}}~\cite{Alwall:2014hca}. We use the on-shell
renormalisation scheme, and have verified UV finiteness both
analytically and numerically for the $gg$ and $q\bar q$ channels
independently. We use real masses throughout this work, but note that
the discrimination of signal and background can have shortfalls
when the scalar width becomes comparable to the resonance
mass~\cite{Nowakowski:1993iu,Seymour:1995np,Passarino:2010qk,Kauer:2015hia,Englert:2015zra},
which is indicative of a loss of perturbative
control~\cite{Goria:2011wa}.

%
\begin{figure*}[!t]
\centering
\includegraphics[width=0.7\textwidth]{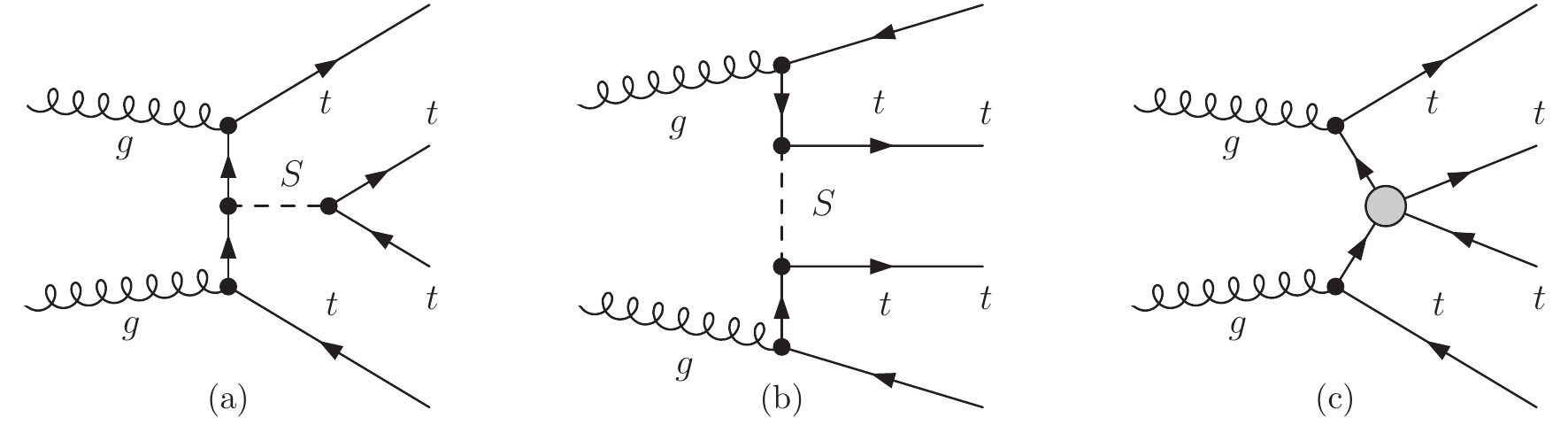}
\caption{\label{fig:fourtop} (a), (b) tree-level graph in the theory of
  eq.\eqref{eq:lagsimp} contributing to four top production; (c)
  tree-level contribution in the EFT description upon integrating out
  the heavy scalar, where the grey blob represents the operator of
  eq.~\eqref{Ott}.}
\end{figure*}

\begin{figure*}[!t]
\centering
\includegraphics[width=0.63\textwidth]{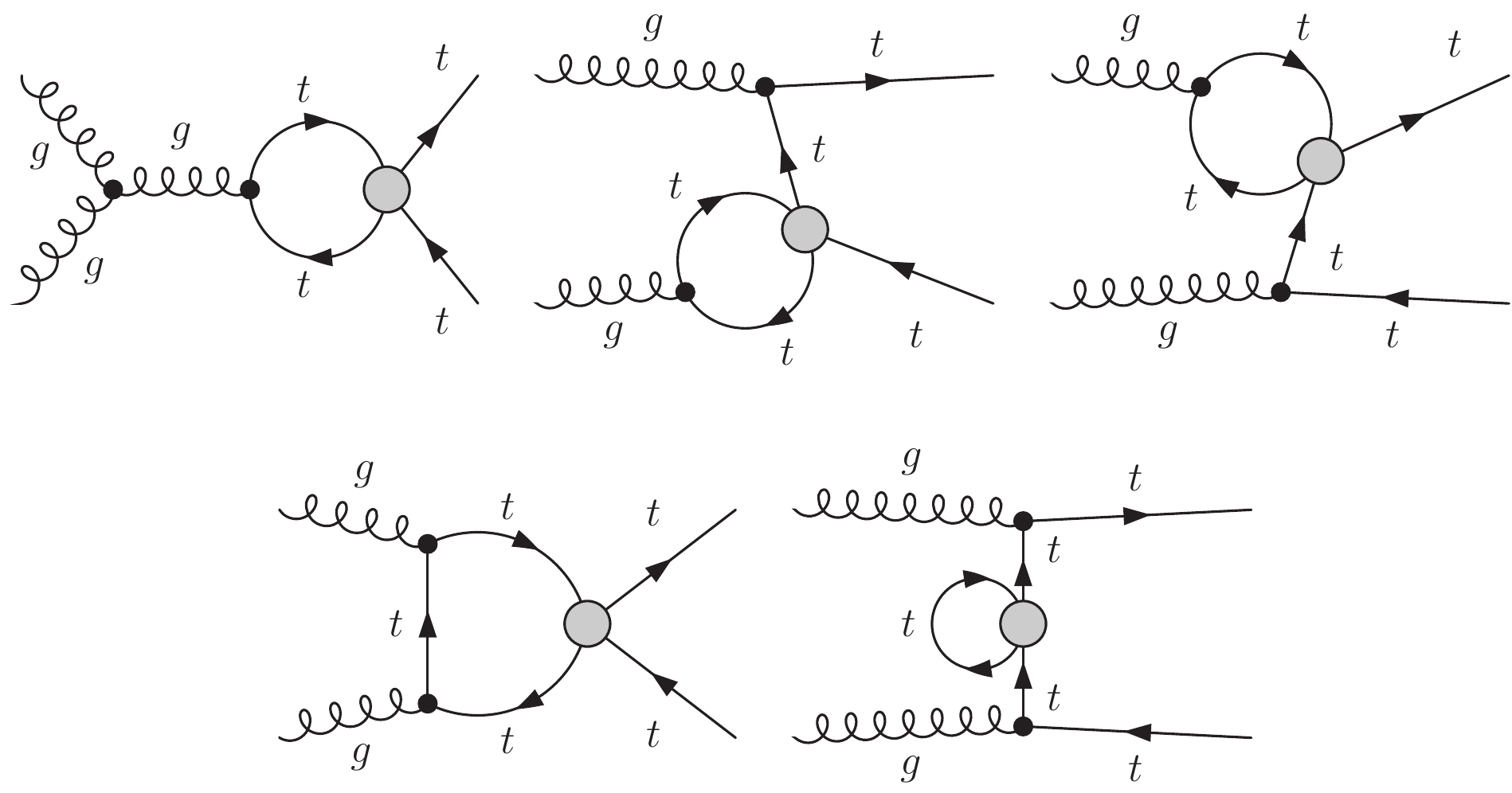}
\caption{Representative one-loop Feynman diagram contributions to $gg\to t\bar t$ arising in the effective theory formulation of eq.~\eqref{eq:lagsimp}, the shaded region represents a four top insertion. \label{fig:effone}}
\end{figure*}

We now turn to the effective theory description of the model of
eq.~\eqref{eq:lagsimp} at low energies or, equivalently, when the
scalar mass $m_S$ is taken to be large. Integrating out the heavy
scalar generates two dimension six operators that enter the processes
considered in this paper. The first of these is a modified
gluon-$t\bar{t}$ interaction, described by the effective operator
\begin{equation}
{\cal{O}}_{tG} = \vev \bar t_L T^a \sigma^{\mu\nu} t_R \, G^{a}_{\mu\nu} 
\label{Otg}
\end{equation}
(and its Hermitian conjugate). Here $t_L$ and $t_R$ denote left-handed and
right-handed top quarks, $T^a$ are the SU(3) generators in the fundamental
representation, \mbox{$\sigma^{\mu\nu}=i[\gamma^{\mu},\gamma^{\nu}]/2$} and
$G^a_{\mu\nu}$ is the QCD gauge field strength tensor. Note, that we have
scaled the operator by an additional factor of the vacuum expectation value
\vev of the SM Higgs boson. The second operator is a four-fermion
operator involving four top quarks, and given by expanding the scalar propagator
for large $m_S$ in relation to its four momentum $q^2$,
\begin{equation}
(\bar{t}t) {c_S^2\over q^2 -m_S^2} (\bar{t} t) \stackrel{q^2\ll m_S^2}\longrightarrow -{c_S^2\over m_S^2} (\bar{t}t)^2 =  {c_{tt}\over \Lambda^2} {\cal O}_{tt}
\label{Ott}
\end{equation}
see e.g. fig.~\ref{fig:fourtop}. Note that this operator is
not part of the Warsaw (SM EFT) basis~\cite{Grzadkowski:2010es}, 
but it is more convenient for our purposes. For instance in four top production, the operator of
eq.~\eqref{Ott} enters at tree-level in the EFT, as illustrated in
fig.~\ref{fig:fourtop}. The contribution from the operator
${\cal{O}}_{tG}$ is suppressed with respect to the contribution from
$\mathcal{O}_{tt}$ because it is loop-induced and four-top contributions with
one ${\cal{O}}_{tG}$ insertion are of higher order in $\alpha_s$ than four-top
contributions with one insertion of ${\cal{O}}_{tt}$ (see appendix~\ref{sec:appen}). The situation is
different in top pair production. Since ${\cal{O}}_{tt}$ enters only through
loops (see, e.g. fig.~\ref{fig:effone}) there is no relative suppression with
respect to ${\cal{O}}_{tG}$ which is also loop-induced. Furthermore, tree-level
diagrams with one ${\cal{O}}_{tG}$ insertion (whose topology is the same as the
upper three diagrams in fig.~\ref{fig:effct}) and one-loop diagrams with one
${\cal{O}}_{tt}$ insertion (as in fig.~\ref{fig:effone}) contribute to the same
perturbative order in $\alpha_s$.  Hence, in top quark pair production both
operators are relevant and have to be included in a consistent EFT calculation.

\begin{figure}[!t]
\centering
\includegraphics[width=0.48\textwidth]{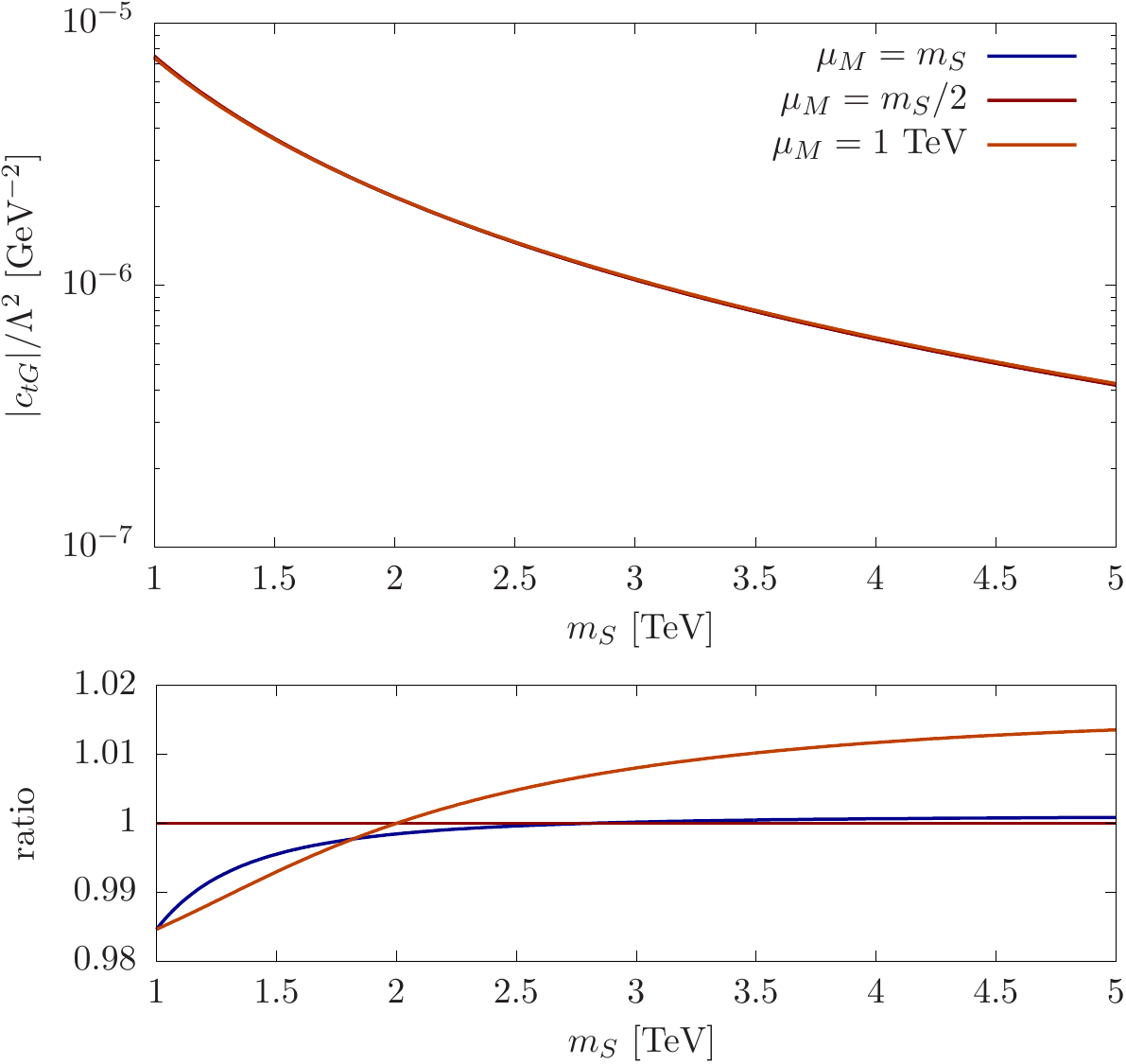}
\caption{
Matched value of $c_{tG}$ for different matching scale choices and scalar masses ($c_S=1$) as detailed in the text. \label{fig:vertexcomp}
}
\end{figure}

\begin{figure*}[!t]
\centering
\includegraphics[width=0.48\textwidth]{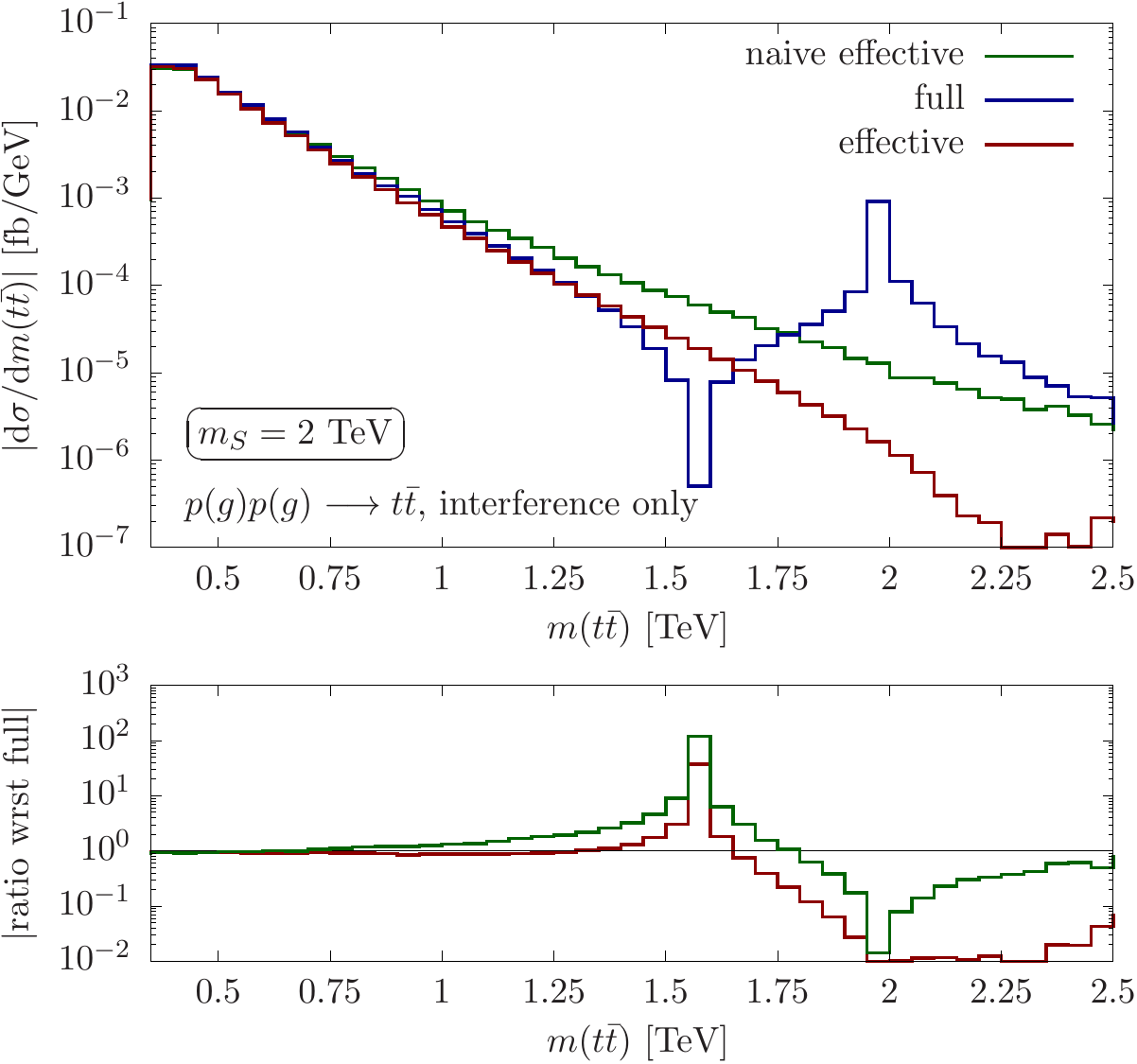}
\hfill
\includegraphics[width=0.48\textwidth]{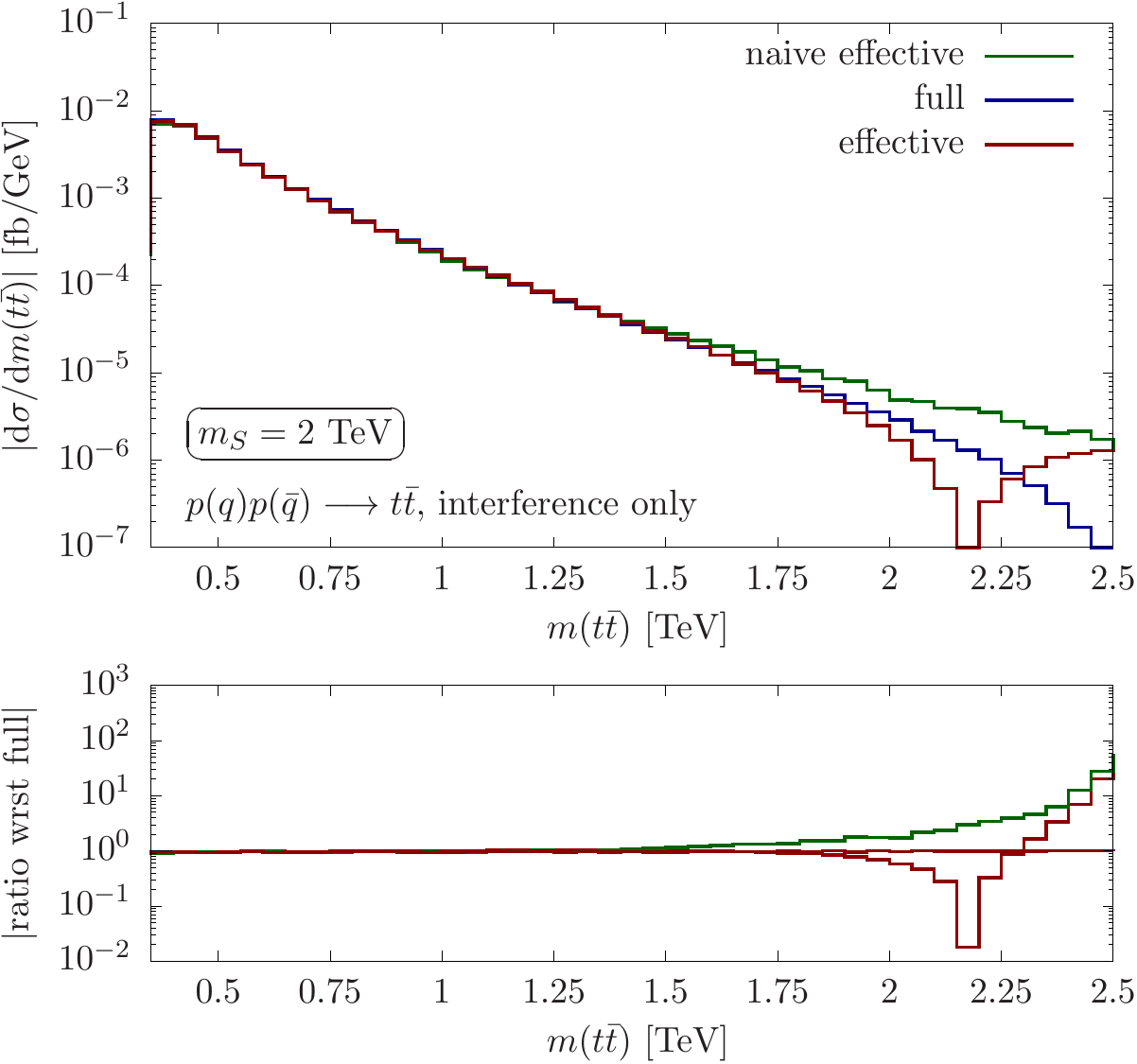}
\caption{BSM interference contribution as a function of the invariant $t\bar t$ mass for gluon fusion (left) and $q\bar q$ annihilation, neglecting the $Z$ contribution. As the interference changes sign we choose to plot the absolute value of the interference for clarity. We choose $m_S=2$~TeV and $c_S=0.1$. \label{fig:eftmatch}}
\end{figure*}

In particular, including the four top contribution requires the calculation
of the loop diagrams of fig.~\ref{fig:effone}.
These contributions require additional
UV counterterms which are not present in the full theory but which
renormalise the EFT operators.  For illustrative purposes, we show the
counterterm graphs in the gluon channel in
figure~\ref{fig:effct}. Renormalisation of the UV divergences related to EFT operators is
performed in the $\overline{\text{MS}}$ scheme, and we have checked UV
finiteness of all of our expressions for the final amplitude.
The sum of tree-level contributions from ${\cal{O}}_{tG}$ and the one-loop
contributions from ${\cal{O}}_{tt}$ yields the results for observables of top
quark pair production at NLO EFT in terms of the Wilson coefficients $c_{tG}$ and $c_{tt}$.

Furthermore, our use of a
specific model of new physics at high energy means that we can fix
the values of the Wilson coefficients by matching the full theory and
NLO EFT calculations at a suitable matching scale $\mu_M$ taking the operator
mixing between ${\cal{O}}_{tG}$ and ${\cal{O}}_{tt}$ into account. We choose $\mu_M=m_S/2$ in the
following unless otherwise specified. We extract $c_{tG}$ as the
finite remainder after subtracting the
$\overline{\text{MS}}$-renormalised four fermion one-loop insertion
from the EFT operator that is induced by the propagating $S$. Note
that it does not require UV renormalisation as opposed to the four
fermion insertion. The dependence of $c_{tG}$ on the matching scale is
shown in figure~\ref{fig:vertexcomp}. As the matching scale is related
to a renormalisation scale choice (appendix~\ref{sec:appen}), the cross section
has a logarithmic dependence on the $\mu_M$.

As for the full simplified model calculation described above, we have
implemented our matched NLO calculation in a modified version of
{\sc{Vbfnlo}}~\cite{Arnold:2008rz,Arnold:2011wj,Arnold:2012xn,Baglio:2014uba}.

\begin{figure*}[!t]
\centering
\includegraphics[width=0.48\textwidth]{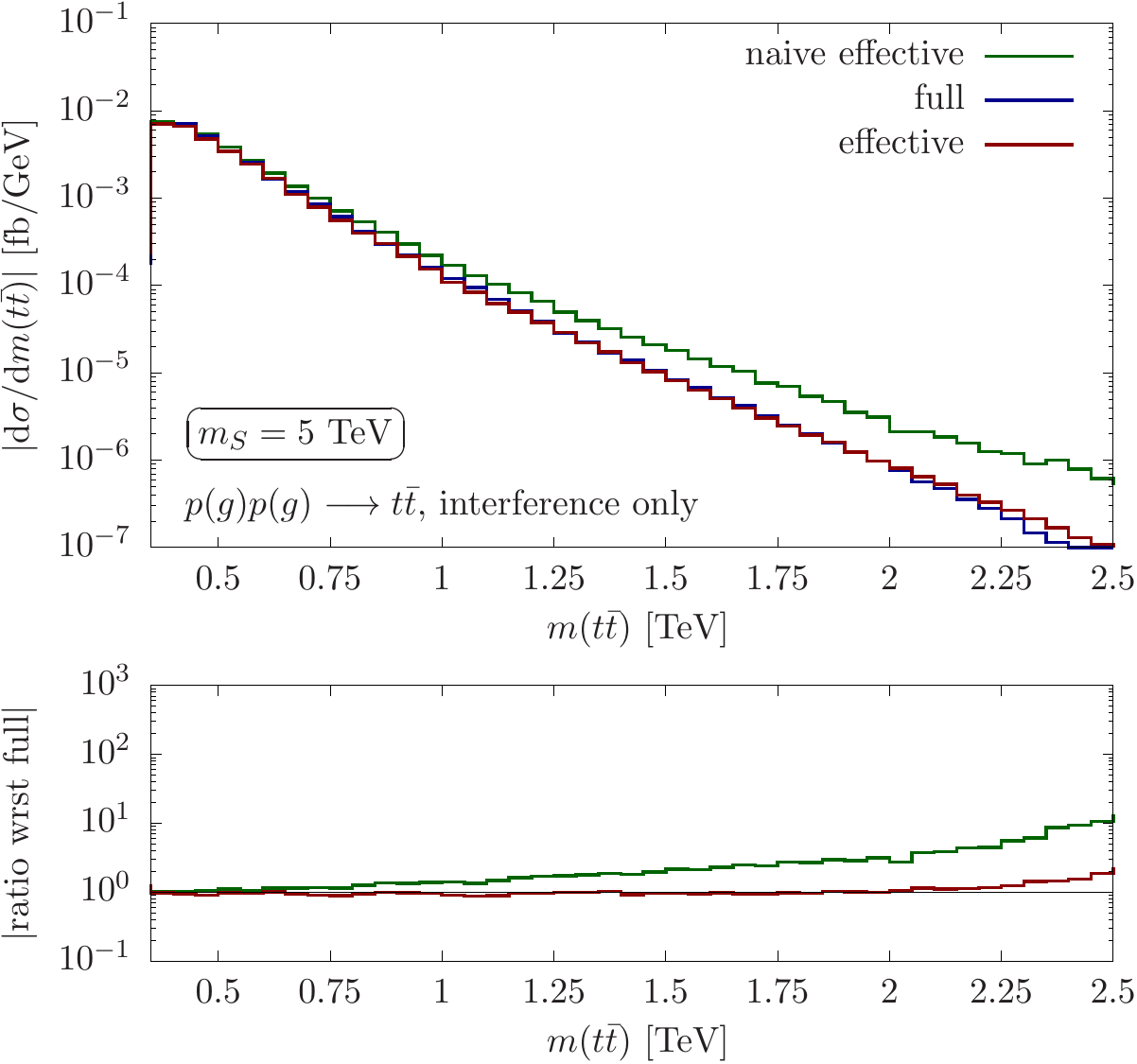}
\hfill
\includegraphics[width=0.48\textwidth]{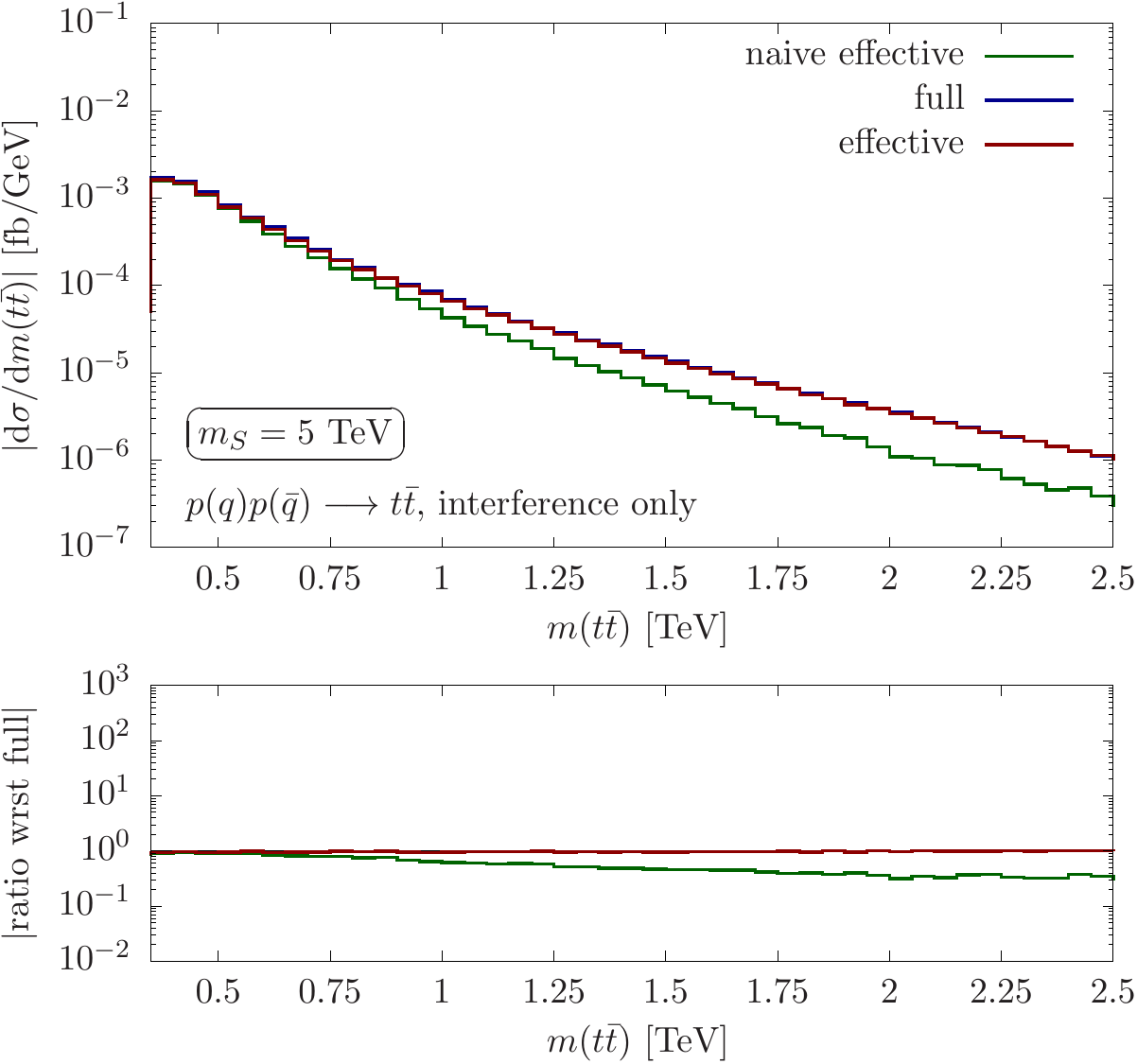}
\caption{Like fig.~\ref{fig:eftmatch} but choosing $m_S=5$~TeV and again $c_S=0.1$ to demonstrate the broad range
of agreement for heavy scalar masses probing the LHC kinematical coverage.\label{fig:eftmatch2}}
\end{figure*}

\section{Validity of EFT at (next-to) leading order}
\label{sec:validity}

In the previous section, we outlined a particular simplified model for
new physics in the top quark sector, and explained how this can be
matched to an EFT description at low energies. In this section, we
analyse the range of validity of the latter, as the mass of the scalar
particle is lowered towards LHC energies. We will illustrate our
results using the invariant mass distribution of the final state tops
in top pair production, although similar results would be obtained for
other kinematic distributions. 

In fig.~\ref{fig:eftmatch}, we show the contribution to the
invariant mass $m_{t\bar{t}}$ stemming from the interference between
the new physics process, and the SM contribution, 
\begin{equation}
\text{d}\sigma  (t\bar t)\sim 2\text{Re}\left( {\cal{M}}_{t\bar t}^{\text{SM}} {\cal{M}}_{t\bar t}^{\ast\,\text{virt/d6}}\right)
\end{equation} 
where virt/d6 represents the propagating-$S$ contributions or their
dimension six EFT counterparts, for a scalar mass of $m_S=2$ TeV.  Three
different curves are shown.  The blue curve shows the result obtained
from the full theory of eq.~(\ref{eq:lagsimp}), with all dynamics
correctly included. The red curve shows the results of our NLO-matched
EFT calculation. We see that the EFT and the full computation
agree well, as long as we are away from the turn-on of the scalar
Breit-Wigner distribution. The green curve in fig.~\ref{fig:eftmatch}
shows the results of a bottom-up approach to EFT where we assume no knowledge
about the full theory. Specifically, we perform a LO EFT calculation of $t \bar
t$ production taking only tree-level diagrams with one insertion of
${\cal{O}}_{tG}$ into account. We treat the Wilson coefficient $c_{tG}$ as a
free parameter in the EFT and fit $c_{tG}$ to Monte Carlo data that was
generated using the full theory.  This approach simulates an EFT fit where the
EFT prediction is calculated at LO and applied to data which contains the
signatures of the simplified model of eq.~\eqref{eq:lagsimp}.
This na\"ive approach based on fitting $c_{tG}$
alone never reproduces the correct shape.  This becomes even more
transparent when we push the scalar mass to larger values,
e.g. $m_S=5$ TeV in fig.~\ref{fig:eftmatch2}.  The full theory and 
the NLO EFT calculation agree very well, with the turn-on of the scalar exchange
only leading to mild corrections for large values of $m(t\bar t)$ in
the (dominant) gluon fusion component. Again as expected, the LO EFT
approach now deviates significantly. In particular, fixing the
coefficient of $c_{tG}$ at low energies where the $m_{t\bar t}$ distribution
is measured more precisely leads to a drastic
mismodelling of the shape of the invariant mass distribution, with a
significant overestimate of the high mass tail. As we will see in the
following section, this can lead to an overly optimistic constraint on
possible new physics effects, for the model that we consider here.

\begin{figure}[!t]
\centering
\includegraphics[width=0.43\textwidth]{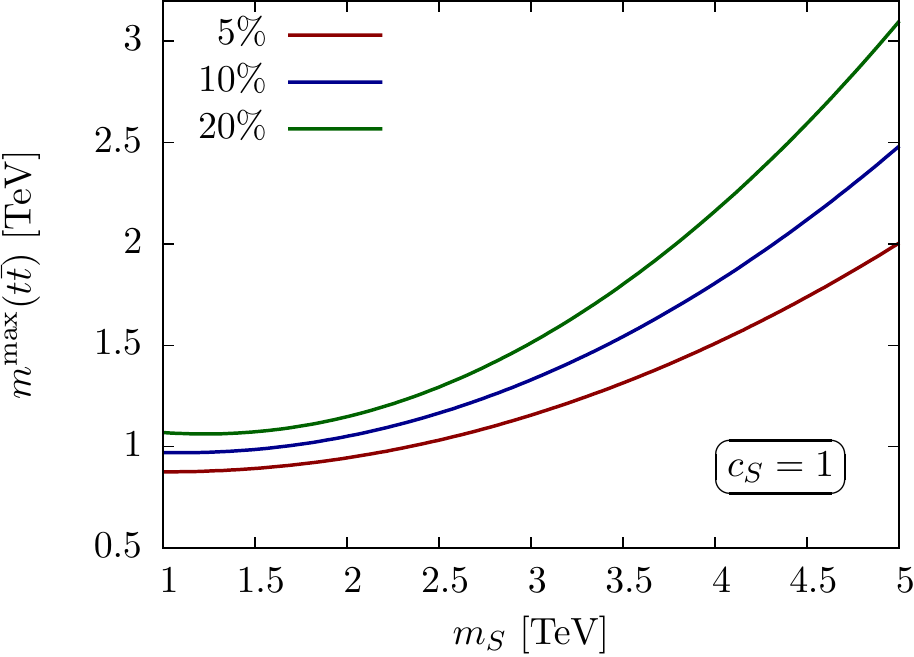}
\caption{Validity within percentage of the EFT computation when compared to the full model including squared Higgs resonance contribution. $m^{\text{max}}(t\bar t)$ indicates the invariant \mbox{(anti-)top} mass, at which the relative difference becomes larger than the indicated percentage.
\label{fig:compare}}
\end{figure}

In fig.~\ref{fig:compare}, we indicate the validity range when
comparing full theory and NLO EFT computation (for a general discussion
see \cite{Contino:2016jqw}). The parameter $m^{\text{max}}(t\bar t)$
denotes the energy scale at which the NLO EFT and full computations
deviate beyond the indicated percentages for $c_S=1$. In this
comparison we also include the squared resonance contribution. Note
that for this coupling choice, the width remains at $\simeq 0.1~m_S$
leading to the turn-on of the Breit-Wigner distribution becoming
resolvable in the direct comparison. This turn-on cannot be resolved
when background uncertainties are included (see below).

Our results in this section confirm the possibility of obtaining an
accurate EFT description of the simplified model of
eq.~(\ref{eq:lagsimp}), which is generic enough to apply in multiple
contexts, including singlet Higgs mixing scenarios and multi-Higgs
doublet extensions. A key issue facing contemporary global EFT fits is
whether or not to pursue the effort of carrying out a full NLO
calculation for all processes and observables considered. The latter
requires a considerable effort (see e.g.~\cite{Hartland:2019bjb,Brivio:2019ius} for
recent examples), although the intermediate possibility also exists of
including renormalisation group mixing effects between dimension six
operators, but neglecting additional contributions that are
non-logarithmic in the matching scale. The importance of NLO effects
in the present case is ultimately due to the fact that of the two
operators that are sourced in the low energy description, one is
tree-induced but the other is loop-induced. Our example thus clearly
shows the need to bear such considerations in mind when trying to
match EFT constraints to specific new physics models.

\section{LHC comparisons, results and extrapolations}
\label{sec:results}

\begin{figure}[!b]
\centering
\includegraphics[width=0.45\textwidth]{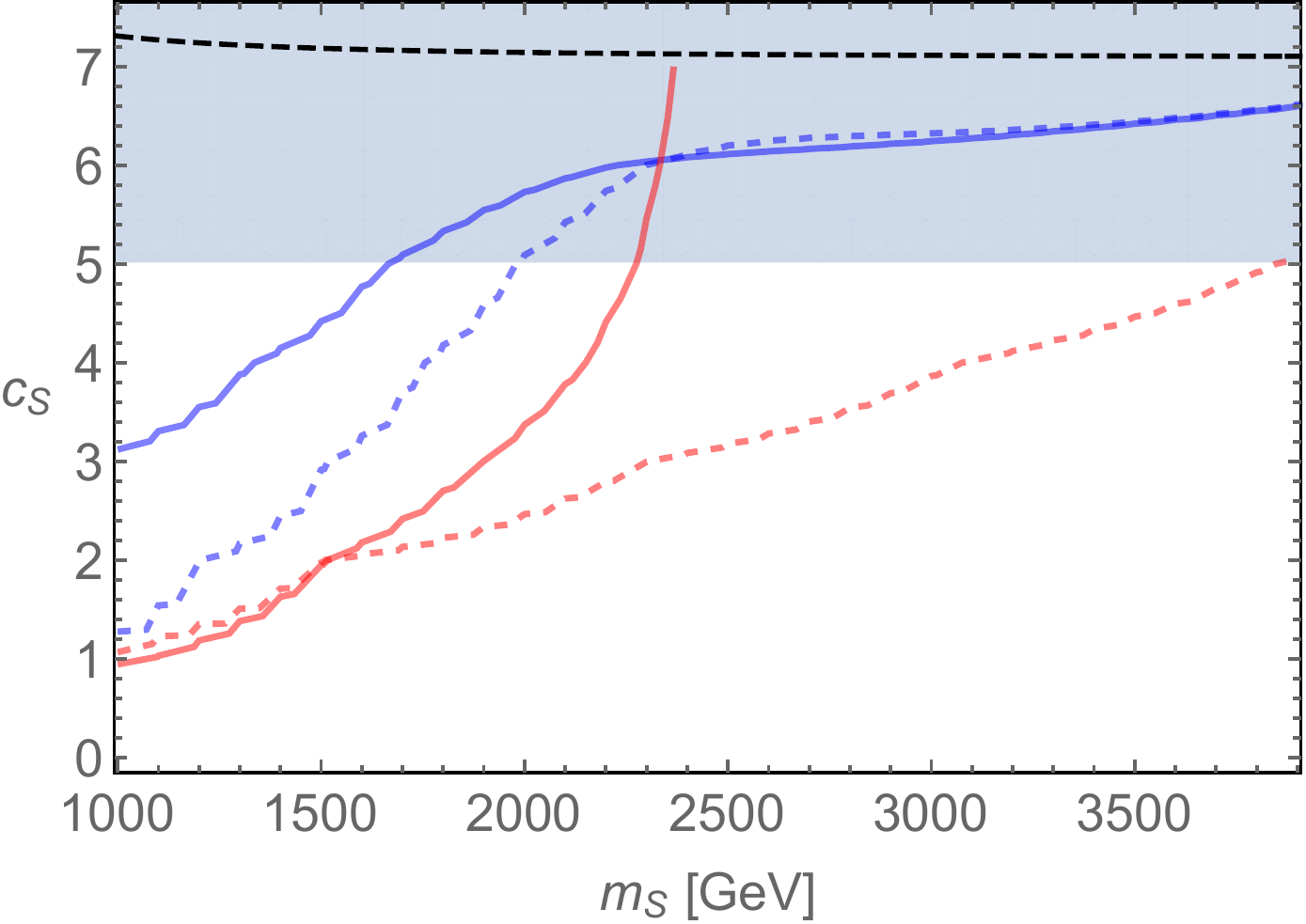}
\caption{95\% confidence level exclusion contours for the simplified model of eq.~\eqref{eq:lagsimp} as a function of its mass $m_S$ and top coupling $c_S$. The blue solid contour shows the full result (i.e. propagating $S$ at NLO) while the blue dashed line corresponds to the EFT calculation. For $pp\to t\bar t$ we assume a flat uncertainty of 3\%. The solid red line represents a $pp\to t\bar t t\bar t$ analysis of the simplified scenario using the extrapolation of Ref.~\cite{ATL-PHYS-PUB-2018-047} while the red dashed line represents the (LO) EFT four top results. The shaded band shows the region
where perturbative unitarity is lost, $c_S\gtrsim \sqrt{8\pi}$ which we obtain from an explicit partial wave projection calculation of $t\bar t \to t\bar t$ in the full model, i.e. with propagating $S$. Note that this is precisely the region where $\Gamma(S\to t\bar t)\simeq m_S$ according to Eq.~\eqref{eq:partialtt}. Finally, the black dashed line is the unitarity constraint on the effective four top interaction, below which unitarity is preserved (for details see text).
\label{fig:confidence}}
\end{figure}

Both top pair and four top production are being actively measured at
the LHC, and will play a crucial role in searching for new physics in
the top quark sector in the coming years. To this end, it is
instructive to examine the sensitivity that the LHC is likely to
achieve following its high luminosity upgrade, with an expected 3
{ab}$^{-1}$ of data. We will do this here for two scenarios. Firstly,
we will constrain the full simplified model of eq.~(\ref{eq:lagsimp})
directly. Assuming a given uncertainty for the above-mentioned
processes leads to exclusion contours in the $(m_S, c_S)$ parameter
space, shown in fig.~\ref{fig:confidence}, where anything above a
given curve (i.e. for stronger couplings $c_S$) is excluded. Secondly,
we will assume that an NLO EFT analysis has been applied, leading to
constraints on the coefficients of the new physics operators ${\cal
  O}_{tG}$ and ${\cal O}_{tt}$. By matching with the full theory as
described previously, constraints on the operator coefficients can
also be converted to curves in the $(m_S,c_S)$ plane.

The top pair production cross section is currently known at NNLO
precision~\cite{Czakon:2013goa,Czakon:2015owf} (see
also~\cite{Czakon:2018nun}). Given the large cross section, the
theoretical uncertainty will be the limiting factor of physics in the
top sector (see also~\cite{Englert:2016aei}).  In
fig.~\ref{fig:confidence}, we show the sensitivity of the LHC under
the assumption that the unfolded $m_{t\bar t}$ distribution can be described at an optimistic 3\% level
using a binned $\chi^2$ test as detailed in
Ref.~\cite{Buckley:2015lku}. For this particular error choice the EFT
and full theory agreement happens to be slightly above the perturbative unitarity limit
of $c_S^2 \simeq 8\pi$ that can be derived from $t\bar t \to t\bar t$ scattering in the full model (i.e. with propagating $S$).
  A larger
error budget quickly pushes the constraints deeply into the non-perturbative regime. On the
other hand sensitivity to $c_S\simeq 1$ requires per mille level uncertainties.
These are beyond the current state-of-the-art. As can be seen, for large scalar masses where the EFT
reproduces the full model expectations both approaches are compatible.
At lower masses, tighter constraints
are obtained in the EFT than in the full theory. This is due to the
systematic tendency (visible in
figs.~\ref{fig:eftmatch}--\ref{fig:eftmatch2}) of the EFT to
overestimate the full theory due to the absence of absorptive parts in
the region where the scalar contribution gets resolved. Thus, applying EFT
alone would result in overly optimistic reported constraints on new
physics, that would not be strictly valid. Note that in this
comparison we include the squared $s$-channel scalar contribution with
an approximate $K$ factor $\simeq 2.5$ \cite{Anastasiou:2015ema,Anastasiou:2016cez}
as this significantly impacts
the exclusion for the dynamic $S$. Notwithstanding the accuracy at
which the EFT manages to approximate the full computation, we see that
hadron collider systematics do seriously curtail precision physics in
the top sector 
when contrasted with certain classes of top-philic BSM models. The simplified
model highlights this through Fig.~\ref{fig:confidence}. Gaining sensitivity in such an instance crucially rests on more precise SM predictions that allow constraints to be pushed into the perturbative limit of the model.

One might argue that discovering a contrived top-philic new physics
scenario is difficult to achieve in the first place. However, for the
scenario that we have studied there is the possibility to
investigate four top final states similar to existing
analyses~\cite{Cao:2016wib,Zhang:2017mls,Englert:2019zmt}. The
experiments have also performed extrapolations to the HL-LHC,
e.g.~\cite{CMS:1900mtx,ATL-PHYS-PUB-2018-047,Azzi:2019yne}. As the
cross sections for this process are relatively small,
${\cal{O}}(10~\text{fb})$~\cite{Bevilacqua:2012em,Frederix:2017wme},
statistical and experimental uncertainties will be important. There is
reason to believe that the latter can be brought under sufficient
control and e.g. ATLAS have shown that a sensitivity of $11\%$ around
the SM expectation can be achieved~\cite{ATL-PHYS-PUB-2018-047} which
is smaller than the current theoretical precision. It is not
unreasonable to expect that theoretical predictions can be improved
and we assume a 18\% accuracy in the extraction of the unfolded $t\bar
t t\bar t$ cross section, which is slightly worse than the ATLAS
extrapolation and the lowest bound provided by CMS~\cite{CMS:1900mtx}.

We simulate four top events using MadEvent~\cite{Alwall:2014hca}
keeping track of destructive interference effects that arise between
the QCD and new scalar contributions. In the four top case, these are
much smaller than for $gg\to t \bar t$, we find a typical mild
correction of ${\cal{O}}(-10\%)$. Constraints on the parameter space
from applying the full simplified model, and the EFT approach, are
shown in fig.~\ref{fig:confidence}. Given that there is a tree-level
induced dimension six operator in this process (i.e. the four-fermion
operator), we restrict the present discussion to LO only. For low
scalar masses the constraints are comparable. However, for larger
masses applying the full model directly leads to very weak
constraints. This behaviour is dominated by the assumed uncertainties,
coupled with the fact that at higher masses in the full theory, the
decay width of the scalar (from
eq.~(\ref{eq:partialtt})) becomes large. This
decreases the scalar contribution to four top final states to a large
extent, leading to a loss of sensitivity for the simplified model in
four top final states under our assumptions at around $m_S\simeq
2.2~\text{TeV}$. The Breit-Wigner cross section distribution of the
$t\bar t$ system for large enough $c_S$ behaves as $\sim
\tilde{\Gamma}^{-2}$ (see eq.~\eqref{eq:partialtt}), i.e. flattens out
as a function of $c_S$ such that the cross section constraint for
large enough $c_S$ is determined by the mass $m_S$. This means that
the sensitivity translated to our simplified model calculation is no
longer under perturbative control.\footnote{The width being related
to the resummation of the imaginary part of the top 2 point function
signifies the relevance of higher order corrections in this model
for the four top final states as well. These effects would be interesting
to study but are beyond the scope of this work.}  
This effect is mainly driven by the dynamics in the full theory
of the chosen model, and is not a problem of setting constraints using EFT approaches.
Hence, applying a cut on the typical energy scale of the process as
proposed for example in \cite{Contino:2016jqw} would not resolve this problem.
While the high mass region is plagued by large width effects in the full theory
the lower masses ($< 2$ TeV) shown in Fig.~\ref{fig:confidence} are accessible 
at the LHC. Hence, matching to the full scenario is not possible in a perturbatively meaningful 
way as mentioned above and illustrated by the discrepancy between the blue solid and dashed
lines in Fig.~\ref{fig:confidence} in the low mass region. The relatively good agreement
between the red solid and dashed lines in that mass region is accidental.

However, without considering a specific UV model we can check the self-consistency
of the EFT with respect to perturbative unitarity. Such constraints
arise from four top scattering in the EFT, i.e. through the four-top contact interaction
in \Eq{Ott} and are given by
\begin{equation}
c_S^2 \lesssim 16\pi {m_S^2\over {E}^2_\text{max} - 2 m_t^2}\,,
\end{equation}
where, for illustrative purposes, we have matched the constraint on $c_{tt}$ to $c_S$ using \Eq{Ott}.
$E_\text{max}$ denotes the energy where unitarity is broken
in the leading order approximation. This happens at the latest at $E_\text{max}\simeq m_S$
when considering our scenario, which is depicted by the black dashed line
in Fig.~\ref{fig:confidence}. Comparing this bound with the red dashed curve shows that the EFT constraints on $c_S$
(or rather $c_{tt}$) from four-top final states remain within the perturbative
unitarity bounds of the EFT. Hence, analyses of the four-top final states
do give rise to constraints which are perturbatively meaningful in 
different scenarios where matching is possible~(e.g.~\cite{Greiner:2014qna,Baek:2016lnv,Fox:2018ldq}).

\section{Summary and Conclusions}
\label{sec:discussion}
Effective field theory approaches are becoming a new standard for the
dissemination of LHC physics results. In contrast to flavour physics
where EFT methods have been successfully employed over decades (see
e.g.~\cite{Buchalla:1995vs}), the non-obvious scale separation of
hadron collider measurements that probe a broad partonic centre of
mass energy range makes their implementation less straightforward. In
particular, operator mixing effects that are sensitive to whether
dimension six operators are tree- or loop-induced in particular UV
scenarios will shape the phenomenology at intermediate scales and has
to be reflected consistently in any limit setting procedure. In this
work we have examined a particular scalar simplified model with
top-philic couplings that approximates a broad range of UV scenarios,
with the particular aim to gauge the sensitivity reach of top quark
final states at the LHC. Top pair production processes with large
cross sections are prime candidates to look for new physics effects
with statistical control.  We demonstrate that the NLO matching of the
EFT and full model allows a broad range of agreement of the two
approaches, up to $\sim 3$ TeV (in e.g. the top pair invariant mass
distribution) for order one coupling choices of the simplified
model. This agreement can be pushed higher when couplings are such as
to reduce the Breit-Wigner-induced departure for invariant masses
below the resonance threshold. For our simplified scenario we find
that systematic limitations of top pair measurements dilute the
sensitivity to new physics effects in particular in the region where
the EFT approach (which is non-trivial in this context) provides an
excellent approximation to the UV theory. In this sense, repeating the
Higgs discovery success story where precision effects allowed to
constrain the Higgs' presence marginally outside LEP's kinematic
coverage seems unlikely when $pp\to t\bar t$ is considered in
isolation (i.e. no other competing BSM effects are present). Pivotal
to changing this situation is the continued precision calculation
efforts for SM processes, and $t\bar t$ production in particular in
our context. In the concrete case of top-philic interactions as
expressed by the scalar model, subsidiary measurements such as four
top final states can provide additional sensitivity. While these
processes are considerably more rare than top pair production at the
LHC, they have direct sensitivity to four top contact interactions
which are clear signs of top-philic interactions below their
characteristic scale. Including four top final states in leading order
fits is therefore crucial to achieve sensitivity to the scenario
discussed in this work, as an example for new physics that
predominantly talks to the top sector.

\acknowledgements 
This work was supported by the Munich Institute
for Astro- and Particle Physics (MIAPP) of the DFG Excellence Cluster
Origins (\url{www.origins-cluster.de}). CE and PG are supported by the
UK Science and Technology Facilities Council (STFC), under grant
ST/P000746/1. CDW is supported by the STFC, under grant ST/P000754/1,
and by the European Union Horizon 2020 research and innovation
programme under the Marie Sk\l{}odowska-Curie grant agreement
No. 764850 ``SAGEX''.

\appendix

\section{Notes on renormalisation and matching}
\label{sec:appen}
\renewcommand{\Re}{\hbox{Re}}
The UV divergent corrections of top pair production in the simplified model are given by the vertex and propagator corrections depicted in Fig.~\ref{fig:full}. The on-shell renormalisation of UV divergencies is determined only
by top quark mass and wave function counterterms (these can be found in Ref.~\cite{Denner:2006fy}).
The cancellation of UV singularities along these lines is expected by the gauge-singlet character of $S$ and the product-group 
gauge theory form of the SM. Hence, there is no renormalization of the gauge couplings.

The qualitative changes in the renormalisation procedure when 
comparing full and effective theory computation is highlighted by considering the top quark two-point function.
Approaching the limit $m_S\to \infty$ before carrying out the loop integration results in a schematic identification
\begin{multline}
\label{eq:expand}
B_0(q^2, m_t, m_S) \sim 
\parbox{3cm}{\vspace{0.5cm}\includegraphics[width=3cm]{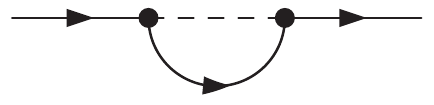}}\\
\stackrel{m_S\to \infty}{\longrightarrow}
-{1\over m_S^2}
\parbox{3cm}{\vspace{0.55cm}\includegraphics[width=3cm]{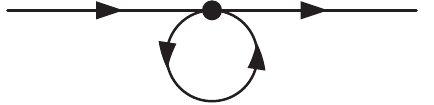}}
\sim - {A_0(m_t)\over m_S^2}\,,
\end{multline}
where $A_0$ and $B_0$ are the Passarino-Veltman one-point and two-point scalar functions~\cite{Passarino:1978jh,Denner:2006fy}. Since the $A_0$ function does not depend on the momentum of the two-point function there is no top quark wave function renormalisation involved in the EFT calculation. Instead the renormalisation of the EFT calculation is performed in the top quark mass and the Wilson coefficient $c_{tG}$. The EFT renormalisation of the top mass due to the four fermion insertion is given by
\begin{equation}
\delta m^{\text{EFT}}_t = { c_{tt} \over 16\pi^2 \Lambda^2 } m_t A_0(m_t)\,.
\end{equation}
The one-loop EFT contributions (see fig.~\ref{fig:effone}) give rise to UV singularities. After top mass renormalisation we are left with the following UV divergence in the NLO EFT amplitude
\begin{equation}
{\cal{M}}(gg\to t\bar t)\big|_\text{NLO, div}^{\text{EFT}, m_t\text{-ren.}} = -{c_{tt} g_s y_t\over 32\pi^2 \Lambda^2} \Delta^{\text{UV}} \langle {\cal{O}}_{tG} \rangle\,,
\label{eq:ctgdiv}
\end{equation}
where $\Delta^{\text{UV}}= \epsilon^{-1} - \gamma_E + \log 4\pi$ in dimensional regularisation with $D=4-2\epsilon$ dimensions and $y_t$ denotes the top Yukawa coupling (we have traded $m_t$ against the vacuum expectation value that apears in the normalisation of eq.~\eqref{Otg}). 
The amplitude $\langle {\cal{O}}_{tG} \rangle$ denotes all ${\cal{O}}_{tG}$ operator insertions that contribute to $gg \to t\bar t$ at tree-level including those with contact interactions $ggt\bar t$.
This shows that the one-loop insertion of the four-fermion operator ${\cal{O}}_{tt}$ induces a renormalisation of the ${\cal{O}}_{tG}$ operator since the LO EFT amplitude is given by
\begin{equation}
{\cal{M}}(gg \to t\bar t)\big|^{\text{EFT}}_\text{LO} =  \langle {\cal{O}}_{SM} \rangle + {c_{tG}\over \Lambda^2} \langle {\cal{O}}_{tG} \rangle\,,
\end{equation}
where $\langle {\cal{O}}_{SM} \rangle$ represents the SM amplitude, which is independent from $\langle {\cal{O}}_{tG} \rangle$ as a result of \cite{Grzadkowski:2010es}. The divergence in eq.~\eqref{eq:ctgdiv} can be removed by including a $c_{tG}$ counter term
\begin{equation}
\label{eq:ctg_renorm}
{\delta c_{tG} \over \Lambda^2} =  {c_{tt} g_s y_t\over 32 \pi^2 \Lambda^2} \left(\Delta^{\text{UV}} + {\cal{F}}(\mu^2) \right)\,,
\end{equation}
where $\cal{F}$ denotes renormalisation-scheme dependent finite terms that will be fixed when we match the one-loop EFT amplitude with the on-shell renormalised one-loop result for propagating $S$ at a matching scale $\mu_M$. The matching relation (which also addresses the quark-induced channels) is given by
\begin{widetext}
\begin{equation}
\label{eq:ident}
\parbox{16cm}{\includegraphics[width=16cm]{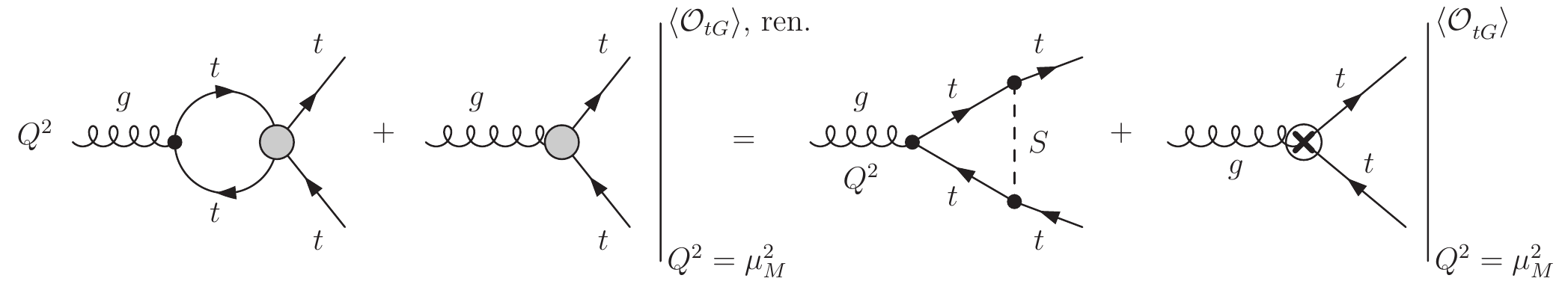}}\,.
\end{equation}
Concretely this means that we first extract the Lorentz structure related to the operator insertion of ${\cal{O}}_{tg}$ of the renormalised EFT as well as the full calculation. We then identify the coefficients of the ${\cal{O}}_{tg}$ amplitudes (Lorentz structures) at a matching scale $\mu_M^2$, which fixes the finite terms ${\cal{F}}(\mu_M^2)$ that correspond to a tree-level insertion of ${\cal{O}}_{tG}$ after matching
\begin{multline}
 {\cal{F}}(\mu_M^2) = -\frac{m_S^2}{\mu_M^2-4 m_t^2} 
+\frac{m_S^2 }{\mu_M^2
  m_t^2-4 m_t^4}\tilde{A}_0(m_S^2) -\frac{m_S^2 }{m_t^2
  (\mu_M^2-4 m_t^2)}\tilde{A}_0(m_t^2)\\+
\frac{m_S^2 (4 m_t^2 (\mu_M^2-4 m_t^2)-m_S^2 (\mu_M^2-10
  m_t^2))}{m_t^2
  (\mu_M^2-4 m_t^2)^2}  \tilde{B}_0(m_t^2,m_S^2,m_t^2) +\left(1-\frac{3 m_S^2 (2 m_S^2+\mu_M^2-4
  m_t^2)}{(\mu_M^2-4 m_t^2)^2}\right)
  \tilde{B}_0(\mu_M^2,m_t^2,m_t^2) \\ -\frac{6 m_S^4
  (m_S^2+\mu_M^2-4 m_t^2)}{
  (\mu_M^2-4 m_t^2)^2} \tilde{C}_0(m_t^2,\mu_M^2,m_t^2,m_S^2,m_t^2,m_t^2)\,,
\end{multline}
\end{widetext}
where $\tilde{A}_0,\tilde{B}_0,\tilde{C}_0$ are the $\epsilon\rightarrow 0$ finite remainders of the Passarino-Veltman one-, two-, and three-point scalar integrals, respectively, in the convention of \cite{Hahn:1999mt,Hahn:2001rv}. This gives rise to a matched value
\begin{equation}
{c_{tG}(\mu_M^2) \over \Lambda^2} = -{c_{S}^2 g_s y_t\over 32 \pi^2 m_S^2}  {\cal{F}}(\mu_M^2) \,,
\end{equation}
which reflects the $g_s$-loop induced nature of $c_{tG}$ in the considered simplified model. The inclusion of appropriately defined finite terms in the comparison of Sec.~\ref{sec:simp} is crucial to obtain agreement between full and EFT-based computation. This balances the $c_{tG}$-related momentum transfer behaviour of $\langle {\cal{O}}_{tG}\rangle$ against the virtual contributions of ${\cal{O}}_{tt}$. In a na\"{i}ve or bottom-up approach based on $\langle {\cal{O}}_{tG}\rangle$ alone without matching this balance is lost which leads to overestimates in the tails of distributions long before $\sqrt{s}\simeq m_S$.

Note that only the sum of loop-inserted $c_{tt}$ and tree-level $c_{tG}$ is defined as a consequence of eq.~\eqref{eq:ident} and we can always move finite terms between the EFT coefficients. The scheme that we adopt is fixing $c_{tt}$ through leading order $t\bar t \to t\bar t$ scattering (accessible in four top final states), which leaves $c_{tG}$ determined as a function of $\mu_M$.

\bibliography{paper.bbl}

\end{document}